\documentclass[11pt,preprint]{aastex}

\newcommand\mdot{\dot{M}}
\newcommand\msun{M_{\odot}}

\newcommand\msunyr{M_{\odot}\, \rm yr^{-1}}
\newcommand\be{\begin{equation}}
\newcommand\en{\end{equation}}

\newcommand\kms{{\, km \, s^{-1}}}

\newcommand\etal{{\rm et al}.\ }

\shorttitle{Why Accretion?}
\shortauthors{Hartmann, D'Alessio, Calvet, Muzerolle}

\begin{document}

\title{Why Do T Tauri Disks Accrete?}

\author{Lee Hartmann\altaffilmark{1}, 
Paola D'Alessio\altaffilmark{2}, Nuria Calvet\altaffilmark{1},
James Muzerolle\altaffilmark{3}}
\altaffiltext{1} {Department of Astronomy, University of Michigan, 
500 Church St., 830 Dennison, Ann Arbor, MI 48109; lhartm@umich.edu, ncalvet@umich.edu }
\altaffiltext{2} {Centro de Radioastronomia y Astrofisica, 
UNAM, Apartado Postal 3-72 (Xangari), 58089 Morelia, Michoacan, Mexico;
p.dalessio@astrosmo.unam.mx}
\altaffiltext{3} {Steward Observatory, University of Arizona, Tucson, AZ 85712; 
jamesm@as.arizona.edu}

\email{lhartm@umich.edu}

\begin{abstract}
Observations of T Tauri stars and young brown dwarfs suggest that the accretion rates of their
disks scale strongly with the central stellar mass, approximately $\mdot \propto M_*^2$.
No dependence of accretion rate on stellar mass is predicted by 
the simplest version of the layered disk model of Gammie
(1996), in which non-thermal ionization of upper disk layers allows accretion to occur
via the magnetorotational instability.  We show that a minor modification of Gammie's
model to include heating by irradiation from the central star yields a modest dependence
of $\mdot$ upon the mass of the central star.  A purely viscous disk model could provide
a strong dependence of accretion rate on stellar mass if the initial disk radius
(before much viscous evolution has occurred) has a strong dependence on stellar mass.
However, it is far from clear that at least the most massive
pre-main sequence disks can be totally magnetically activated by X-rays or cosmic
rays.  We suggest that a combination of effects are responsible for the observed
dependence, with the lowest-mass stars having the lowest mass disks, which can
be thoroughly magnetically active, while the higher-mass stars have higher mass
disks which have layered accretion and relatively inactive or ``dead'' central
zones at some radii.  In such dead zones, we suggest that gravitational
instabilities may play a role in allowing accretion to proceed.  In this connection,
we emphasize the uncertainty in disk masses derived from dust emission, and argue
that T Tauri disk masses have been systematically underestimated by conventional
analyses.  Further study
of accretion rates, especially in the lowest-mass stars, would help to clarify
the mechanisms of accretion in T Tauri stars.
\end{abstract}

\keywords{accretion disks - infrared: stars - stars: formation - stars: pre-main sequence}

\section{Introduction}

The low-mass, pre-main sequence classical T Tauri stars (CTTS) are generally
thought to be accreting mass from their circumstellar disks.  
Disk accretion through stellar magnetospheres  
provides a natural explanation of the ultraviolet and optical
continuum excesses (e.g., Bertout, Basri, \& Bouvier 1988; K\"onigl 1991; Hartigan
\etal 1990; Valenti, Basri, \& Johns 1993; Calvet \& Gullbring 1998) 
and the emission line profiles and strengths of many features 
(Edwards \etal 1994; Hartmann, Hewett, \& Calvet 1994; Muzerolle et al. 2001).

Our understanding of accretion in ionized disks has been revolutionized by
the rediscovery and application of the magnetorotational instability (MRI)
to disk angular momentum transport (e.g., Balbus \& Hawley 1998, and references
therein).  However, it was recognized some time ago that T Tauri disks are cold
and thus likely to have such low ionization levels that 
the MRI cannot operate, at least in some
(and quite possibly extended) regions of the disk,
(e.g., Reyes-Ruiz \& Stepinski 1995).  
This recognition lead Gammie (1996) to suggest that non-thermal
ionization by cosmic rays could lead to accretion in upper disk layers, with
a ``dead zone'' (a non-accreting, non-turbulent region) in the midplane.  Further 
work suggested that X-rays from the central star could also provide a source of
ionization (Glassgold, Najita, \& Igea 1997), and these might even dominate if
low-energy cosmic rays are excluded by turbulent
magnetic fields in the outflows of T Tauri stars.  

While it seems very likely that the very outermost layers of T Tauri disks
are ionized, it is not at all clear that these ionized layers contain enough material
to sustain observed T Tauri accretion rates; 
the column densities of MRI-active material can be quite small 
if dust grains do not settle and/or coagulate to an appreciable extent 
(e.g., Sano et al. 2000).  Beyond this, as we discuss in the following
sections, the standard layered accretion model has difficulties in explaining the 
dependence of mass accretion rates on stellar mass and age.

The other widely-recognized mechanism for transporting mass and angular momentum in disks
that seems viable in pre-main sequence stars is gravitational instability
(e.g., Tomley, Cassen, \& Steiman-Cameron 1991;
Laughlin \& Bodenheimer 1994; Laughlin, Korchagin, \& Adams 1998; see
Gammie 2001, Johnson \& Gammie 2003, and Durisen \etal 2006 
for recent discussions).  This is an especially
attractive possibility in the earliest stages of stellar evolution, as
it is quite likely that most of the mass of stars 
is accreted through the circumstellar disk, due to the finite (and large) angular
momenta of protostellar clouds.  On the other hand, T Tauri stars
have clearly already accreted most of their mass, and typical disk mass estimates
are an order of magnitude below the values required for gravitational
instability (e.g., Beckwith \etal 1990; Andrews \& Williams 2005).
In addition, the surface densities required for gravitational instability
appear to be implausibly large in the innermost disk; some other mechanism
of transport must operate there.

In this paper we discuss the current evidence concerning T Tauri accretion
rates and disk masses and their implications for the mechanisms of mass
and angular momentum transport.  We present a revision of the simple layered model
of Gammie to include irradiation by the central star, which introduces some dependence
upon stellar mass through the dependence of disk heating on the stellar luminosity.  
The inclusion of irradiation does not appear to yield a sufficiently
steep dependence of mass accretion rate on stellar mass to explain all
the observations, although it might be consistent with the upper envelope
of estimated accretion rates.  A fully
viscous disk might be able to explain the observations if the initial
disk size is strongly correlated with the stellar mass, but it is not
clear why disks around at least the most massive stars should be
ionized down through to their midplanes at all radii. 
We suggest that the observed behavior of mass accretion in young stars
may be a complicated mix of layered accretion with possible gravitational
instability in the most massive stars and full viscous accretion in the lowest-mass
objects.

\section{The $\mdot$ - $M_*$ relation}

Observations spanning the range from 2-3 $\msun$ for
intermediate-mass T Tauri stars to brown dwarfs (Calvet \etal 2004; Muzerolle
\etal 2003a, 2005) suggest a strong dependence of mass accretion
rate on stellar mass, roughly $\mdot \propto M_*^2$ (Figure 1).
While observational selection effects tend to produce 
some correlation, because low accretion rates are not detectable 
in higher-luminosity stars, the overall trend of decreasing accretion
rate with decreasing mass is clear (cf. Muzerolle \etal 2005).  
As shown in Figure 1, there is a large scatter in $\mdot$ at a 
given stellar mass $M_*$ among populations of similar age, so that 
mass is not the only parameter controlling accretion. 

One well-known mechanism which produces an $M_*^2$ dependence of
the accretion rate on stellar mass is Bondi-Hoyle accretion:  
\be
{\mdot_{BH}} = { 4 \pi G^2 \rho_{\infty} \over \left (c^2_{\infty} ~+~ v^2_{\infty} \right )^{3/2} }
M_*^2\,, \label{eq:mdotbh}
\en
where $\rho_{\infty}$, $c_{\infty}$ are the gas density and sound speed,
and $v_{\infty}$ is the velocity of the star of mass $M_*$ relative to the ambient gas.
Padoan \etal (2005) noted that the above relation provides the desired
observed dependence of T Tauri accretion on stellar mass, and
argued that the observed mass accretion rates could be explained
quantitatively with realistic values of molecular cloud densities and
velocity dispersions of stars and gas.  Specifically, using moderately plausible values 
of $\rho_{\infty} = 2.46 \, m_H 10^3 {\rm \, cm^{-3}}$,
$v_{\infty} = 1 \kms$, and $c_{\infty} = 0.2 \kms$, equation (\ref{eq:mdotbh}) yields an
accretion rate $\mdot \approx 10^{-8} \msunyr$, consistent with observations
of typical T Tauri stars (Hartmann \etal 1998), although Hartmann (2002) argued
that many of the young stars in Taurus have velocities relative to their natal
gas much smaller than $1 \kms$.

However, using the Bondi-Hoyle result implies that the angular momentum of the 
accreting gas is negligible, so that angular momentum transport in the accretion
disk is unimportant.  This assumption is highly questionable;
if the angular momentum of the star-forming gas is indeed small,
why do many T Tauri stars have such large disks 
($\sim 100$~AU or more, of order $10^4$ stellar radii; e.g., Simon, Dutrey,
\& Guilloteau 2000)?  

Moreover, in Sicilia-Aguilar \etal (2005) we showed that the mass accretion
rates of the T Tauri stars in the $\sim 4$~Myr-old young cluster Trumpler 37
are similar to those of Taurus young stars, even though Tr 37 resides not in
a molecular cloud but within the expanding H II region IC 1396
(see discussion in Patel \etal 1995).  Wendker \& Baars (1980) developed an approximate
model for this inhomogeneous ionized region, based largely on radio continuum observations
at 2.7 GHz.  They estimated an electron density (and thus a proton density) 
$N_e \sim 3.3 \rm{\, cm^{-3}}$ in the central region where most of the cluster stars reside,
although there are nearby denser regions with $N_e \sim 15 \rm{\, cm^{-3}}$.
Even if we take the larger of these values, the ambient gas density in Tr 37 is
two orders of magnitude smaller than the typical value used by Padoan \etal for
Taurus stars.  Moreover, as the cluster resides within an H II region, the sound
speed $\sim 10 \kms$ must dominate the velocity term in equation (1); 
the denominator therefore is three orders of magnitude larger in the case 
of Tr 37 than for Taurus.  This combination of density and velocity factors
results in a predicted Bondi-Hoyle accretion rate five orders of magnitude lower
than observed in Tr 37.  

In summary, Bondi-Hoyle accretion is incapable of explaining the mass accretion
rates in Tr 37, and is unattractive in general because of its great sensitivity
to the properties of the environment.  More broadly, this mechanism completely
ignores the essential problem of angular momentum transport in disks, to which we now turn.  

\section{Layered MRI accretion}

Gammie (1996) pointed out that cosmic rays (or other ionizing radiation
such as X-rays; Glassgold \etal 1997; Gammie 1999)
could produce a sufficient level of ions in ``active layers'' of surface
density $\Sigma_a$ that the magnetic field could couple effectively
to the gas, allowing the MRI to operate.  Assuming that the disk is 
heated primarily by viscous dissipation, Gammie derived an inner disk
accretion rate $\mdot$ for a standard dust opacity law,
\be
\mdot = 1.8 \times 10^{-8} \, \left ( {\alpha \over 10^{-2}} \right )^2
\, \left ( {\Sigma_a \over 100 g \, cm^{-2}} \right )^3 \, \msunyr\,, \label{eq:mdotgam}
\en
where $\alpha$ is the viscosity parameter and the fiducial active layer surface
density is the estimated penetration depth of cosmic rays.
Remarkably, this fiducial value of the mass accretion rate is of the same order
 as the accretion rates seen in typical T Tauri stars of near-solar mass 
(Gullbring \etal 1998; Hartmann \etal 1998; also Valenti \etal 1993).
However, equation \ref{eq:mdotgam} exhibits no dependence upon stellar mass, which
does not agree with the observations. 

Equation \ref{eq:mdotgam} assumes that
irradiation heating of the disk by the central star is not important.
This may not be the case for the
more massive, luminous stars, nor for the brown dwarfs, which have
extremely low mass accretion rates and therefore presumably low viscous
heating.  The situation is mixed for solar-mass T Tauri stars, whose temperature
structure in the innermost disk may or may not be dominated by viscous heating,
depending upon the properties of the dust grains there (e.g., Figure 5a in 
D'Alessio et al. 2001).  We therefore consider the limiting 
case in which the disk heating is
set mainly by the absorption of radiation from the central star.

In Gammie's model, $\mdot$ decreases with decreasing radius.
The mass accretion rate in the inner disk (and therefore the rate onto
the central star) is set by the accretion rate of the layered model 
at the critical radius $R_c$ where the temperature rises to a level
(taken to be 1000 K) sufficient for thermal ionization to activate
the MRI.  In the model with pure viscous heating,
\be
R_c = 0.13 \, \left ( {\alpha \over 10^{-2}} \right )^{2/3}
\, \left ( {\Sigma_a \over 100 g \, cm^{-2}} \right )^{4/3} \, 
\left ( {M_* \over \msun } \right )^{1/3} \,. \label{eq:rc}
\en
In the case where irradiation dominates, the temperature
of the irradiated disk at cylindrical radius R will be 
\be
T^4 \sim {L_* \over 4 \pi \sigma R^2}  \gamma \,, \label{eq:tbb}
\en
where $\gamma$ is a geometrical factor accounting for the angle of entry
of the stellar radiation into the disk.  
We consider two cases.  First, we assume that the critical radius occurs
essentially at the inner edge of the disk where the dust sublimates and 
thus the illumination is at near-normal incidence; this may be reasonable
because the temperatures estimated at this point are $\sim 1400 K$ (Muzerolle 
\etal 2003b), close to the temperature required for thermal activation of the MRI. 
A second case corresponds to the situation when the thermal activation
temperature is achieved
at radii exterior to the dust destruction radius but in the
geometrically flat inner disk, in which case $\gamma \propto 1/R$.

For the first case, thermal activation at the dust destruction radius,
\be
T_c \propto L_*^{1/4} \, R_c^{-1/2}\,
\en
and therefore for fixed $T_c$
\be
R_c \propto L_*^{1/2}\,.
\en
Then for an ``alpha'' viscosity $\nu = \alpha c_s/\Omega$, where
$c_s$ is the central disk sound speed and $\Omega$ is the Keplerian
angular velocity, 
\be
\mdot \propto \nu \Sigma \propto \alpha c_s^2 \Omega^{-1} \Sigma_a 
\en
\be 
\propto \alpha \Sigma_a L_*^{3/4} M_*^{-1/2}\,.
\en
For pre-main sequence stars up to masses of $2 \msun$ or so, the 
stellar luminosity tends to scale very roughly as $L_* \propto M_*^2$.
Therefore in the limit of pure irradiation heating, we would expect
\be
\mdot \propto \alpha \Sigma_a M_*\,. \label{eq:mdotirr}
\en
If instead we assume that the critical temperature is achieved outside
of the dust destruction radius but in the flat disk, then
\be
R_c \propto L_*^{1/3}\,
\en
which results in 
\be
\mdot \propto \alpha \Sigma_a M_*^{1/2}\,. \label{eq:mdotirrflat}
\en

Thus, the inclusion of irradiation heating results in a layered model
in which the accretion rate is no longer independent of stellar mass. 
While the predicted dependence is clearly not as steep as the overall fit to
the observations, it is intriguing that the first model of irradiation
provides a dependence on mass not far from that of
of the upper envelope of the accretion rates seen, as shown by the
solid line in Figure 1, which indicates a relation
$\mdot \propto M_*$.  We return to this point later.  

The detailed calculations of disk models by D'Alessio \etal (1999, 2001)
suggest that irradiation heating does not necessarily dominate viscous heating in the
inner disk, depending upon the mass accretion rate and dust opacity. 
This implies that the above results are limiting cases, and may 
overestimate the dependence of accretion rate on mass in the irradiated
layered accretion model; more detailed models will be required to explore this further.

Equation (\ref{eq:mdotirr}) 
suggests that we look for an additional factor producing a dependence
of either $\alpha$ or $\Sigma_a$ on $M_*$ to improve the layered model.  
If stellar X-rays rather than cosmic rays
are responsible for ionizing the active layer (e.g., Glassgold \etal 1997;
Igea \& Glassgold 1999), the X-ray flux will depend upon the
mass of the young star or brown dwarf, which in principle could
produce a steeper dependence of $\mdot$ on $M_*$.  From the recent study of 
stars in the core of the Orion Nebula Cluster, 
Preibisch \etal (2005) estimated that the ratio $L_x/L_*$ 
is nearly the same for young brown dwarfs as for
young low-mass pre-main sequence stars.  However, the important factor 
is not the total luminosity but the X-ray flux at 
the critical radius where thermal ionization takes over.
In the irradiation limit, the critical radius depends upon the
flux of photospheric radiation.  Since the fluxes of both photospheric
and X-ray radiation should scale in the same way for the same geometry,
there should be no effect.  For example, even though a brown dwarf of mass $\sim 0.08 \msun$
has an X-ray luminosity $\sim 10^{-2}$ that of a typical $0.8 \msun$ T Tauri
star, its $R_c$ will be smaller by the factor $L_*^{1/2} \sim L_X^{1/2}$
in the first case and thus the {\em flux} $F_X (R_c) \propto L_X R_c^{-2}$ will remain 
constant.  The same occurs in the second case of the flat disk, as the geometrical
factors enter in the same way for both the X-rays and photospheric radiation.  

Furthermore,
the calculations of Glassgold \etal (1997) and Igea \& Glassgold (1999) suggest that 
MRI ionization levels are maintained until the
X-rays are very strongly attenuated; this makes the activated total column density 
depend very slowly on the X-ray luminosity.
Thus variations in X-ray irradiation do not seem to produce a
significant dependence upon stellar mass which would steepen the dependence of
mass accretion rate on stellar mass in the layered model.

\section{Viscous (pure MRI) disk evolution?}

The layered model of Gammie (1996) implicitly assumes that the disk has high
enough surface densities that it cannot be magnetically activated all the way
to the midplane by cosmic or X-rays.
If we examine equation (\ref{eq:mdotirr}) from an empirical perspective,
the low accretion rates seen in brown dwarfs -- roughly three orders of magnitude
below typical T Tauri values -- imply much lower disk surface densities in
brown dwarf disks.  This in turn suggests that such low-mass disks might be fully MRI-active.
Indeed, Fromang, Terquem, \& Balbus (2002) suggested that T Tauri disks in 
general might be able to sustain the MRI through their entire extent, even
at reasonably high mass accretion rates, if $\alpha$ is large enough and/or if   
a modest fraction of metal atoms are not locked into grains.

We can examine fully viscous disk evolution schematically by using the toy model
introduced by Hartmann \etal (1998), assuming constant $\alpha$ and the disk
temperature varies as $T \propto R^{-1/2}$.  As shown in that paper, the disk
mass $M_d$ varies with time $t$ as
\be
M_d = {M_d(0) \over (1 + t/t_v)^{1/2}}\,, \label{eq:mdisk}
\en
where $M_d(0)$ is the initial disk mass at time $t=0$, and the 
initial viscous timescale is 
\be
t_v \propto {R_1^2 \Omega \over \alpha c_s^2 }
\propto {M_*^{1/2} \over \alpha T_1} R_1^{1/2}\,,
\en
where $R_1$ is the initial (fiducial) disk radius at $t=0$ and $T_1$ is the 
disk temperature at $R_1$.  With $T_1 \propto L_*^{1/4} R_1^{-1/2}$ and
$L_* \propto M_*^2$, this becomes
\be
t_v \propto {R_1 \over \alpha}\,.
\en

Differentiating equation (\ref{eq:mdisk}) with respect to time to determine 
the mass accretion rate, and taking the limit of significant viscous evolution, 
i.e. $t \gg t_v$, 
\be
\mdot \propto {M_d(0) \over t^{3/2}} { R_1^{1/2} \over \alpha^{1/2}}\,.
\en

It is plausible that the initial disk radius $R_1$ might have a dependence upon stellar
mass.  For example, suppose that the initial protostellar cloud can be approximated
by either a singular isothermal sphere or a Bonnor-Ebert sphere (e.g., Alves, Lada,
\& Lada 2001).  Then the protostellar cloud mass scales with its outer radius as
$M_{out} \propto R_{out}$ for a fixed initial temperature, which we take as relatively
constant.  Then more massive stars will be formed from clouds with larger initial radii.
This has implications for the size of the disk initially formed by the collapse of
the protostellar cloud.  Using the results of the uniformly-rotating 
isothermal sphere collapse calculations of Terebey, Shu, and Cassen (1984),
the material from the outer edge of the cloud falls in to land on the disk at a radius  
\be 
R_d \propto {\Omega_{\circ}^2  R_{out}^4 \over M_*}\,,
\en
where $\Omega_{\circ}$ is the initial angular velocity of the cloud.
Assuming that most of the mass rapidly is accreted into the central mass 
(see \S 7),
i.e. that $M_{out} \sim M_*$, then $R_d \propto \Omega_{\circ}^2 M_*^3$.
If we then identify $R_1 = R_d$ and take $\Omega_{\circ} =$~constant, 
\be
\mdot \propto t^{-3/2} \alpha^{-1/2} M_d(0) M_*^{3/2}\,, 
\en
or, assuming that the initial disk mass scales approximately with the stellar mass
(see \S 7),
\be
\mdot \propto t^{-3/2} \alpha^{-1/2} M_*^{5/2}\,.  \label{eq:mdotvisc}
\en

This purely viscous result yields a strong dependence upon stellar mass, not
too much larger than observed.  (It may be more consistent with the lower
envelope of the data points in Figure 1, but this is uncertain because 
of detection thresholds.)  However, this solution is not without problems.
For example, it is not clear that $\Omega_{\circ}$
should be independent of mass, or that its variation with mass would naturally
produce a result even closer to $M_*^2$.  More importantly, 
low mass accretion rates are accompanied by low disk masses in this
model.  Specifically, the disk mass at long times $t$ scales at a given time as
\be
M_d (t) = M_d(0) {t_v^{1/2} \over t^{3/2}} \propto M_d(0) \left ({ R_1 \over \alpha} 
\right )^{1/2} \propto M_d(0) M_*^{3/2}\,,
\en
i.e. in the same way as the accretion rate.  Thus the pure viscous solution requires
that the brown dwarf disks with very low mass accretion rates have very
low disk masses.  While many brown dwarf disks have undetectable dust
emission (so far), Klein \etal (2003) detected mm-wave emission from two 
young brown dwarfs, and Scholz, Jayawardhana, \& Wood (2006) detected
the 1.3mm emission from 5 of 19 Taurus brown dwarfs (one was previously
detected by Klein \etal).  The interpretation of these data is uncertain
for reasons described more fully in \S 7, but the results
suggest disk/star mass ratios of a few percent in these objects, 
similar to CTTS.  It is not clear that {\em any} brown dwarf disks would
be massive enough to be detected using the simple viscous scaling for low
accretion rates.

It should also be pointed out that the above equations assume substantial
viscous evolution in the disk, which only occurs if the initial disk radius
$R_1$ is sufficiently small or if $\alpha$ is large.  
Collapse to initially large disks would result
in very little overall viscous evolution and make the above analysis inapplicable.
Some of the Hubble Space Telescope images of protostars in Taurus suggest infall
to very large disk radii (Padgett \etal 1999) which would require very large
values of $\alpha$ to result in substantial viscous evolution at typical 
T Tauri ages.  In turn, large values of $\alpha$ needed for faster viscous evolution
are problematic.  Using the toy model of Hartmann \etal (1998) discussed above,
the half-mass disk radius for an $\alpha = 10^{-1}$ would spread to nearly 1500 AU
at an age of 1 Myr; and this may be a lower limit, because it assumes 
viscosity $\nu \propto R$ resulting from a disk temperature $T \propto R^{-1/2}$;
cosmic ray heating will tend to drive the disk temperature to $T \sim$~constant,
in which case the viscosity at large radii will be even higher and the disk
expand even faster.  While it is difficult to assign sizes to T Tauri disks
because of the rapid decrease in surface brightness with radius, 
such large radii and rapid disk evolution seems implausible, especially
given objects like TW Hya, which are still accreting at an age of 10 Myr
(Muzerolle \etal 2000), for which the toy model would
predict a truly enormous disk for $\alpha \gg 10^{-2}$, very much larger than
currently detected.

\section{The dust/metal problem}

The surface density of the active layer in the inner disk is very sensitive
to the presence of small dust particles, which can absorb ions and electrons
in the gas.  The estimates of $\Sigma_a$ of Gammie (1996) for cosmic rays
and Glassgold \etal (1997) and Igea \& Glassgold (1999) for 
stellar X-rays assumed substantial depletion
of the small dust in the upper layers.  While one expects settling and grain
growth to occur naturally in T Tauri disks, it is not clear that the magnitude
of these effects is quantitatively sufficient.

Sano \etal (2000) studied the effects of grain depletion and grain size on the
surface density of the active layer for the case of a fixed cosmic ray ionizing
flux, as in Gammie's original theory.  For a standard interstellar medium grain
size distribution, Sano \etal found that depletion factors of $10^{-4}$ were needed
to produce $\Sigma_a \sim 10^2$ within 1 AU; a depletion of a factor of one hundred
resulted in $\Sigma_a \sim 3 {\rm g \, cm^{-2}}$ at 0.1 AU.  Sano \etal also considered
grain growth in the approximation that all the grains were one (large) size.  For
the case of no depletion, even grain growth to 1 $\mu$m resulted in very small
$\Sigma_a \sim 1 {\rm g \, cm^{-2}}$ at 0.1 AU.  

Empirically estimating the amount of settling and grain growth is difficult from
the current observations, but some rough limits can be set.  D'Alessio \etal (1999)
showed that, for a power-law distribution of grain sizes, maximum limits much greater
than a few microns lead to washing out the $10 \mu$m silicate feature, in disagreement
with observations.  Combining the recent Spitzer IRS survey of 
Taurus disk spectra with disk models (Furlan \etal 2005;
D'Alessio \etal 2006) suggests depletion
factors of order $10^{-1}$ to $10^{-2}$.  These constraints are somewhat
model dependent and uncertain,
but it is difficult to explain the SEDs without a certain amount of disk ``flaring''
which can only be achieved if a significant amount of small dust still remains
suspended in the upper layers.  Thus there is no clear evidence that 
settling and/or grain growth in upper disk layers is large enough to provide
much more than $\Sigma_a \sim 1 - 10 {\rm g \, cm^{-2}}$ at 0.1 AU.  Conversely,
there is evidence for a substantial population of small grains in upper disk
layers, based on the presence of silicate features (D'Alessio \etal 2001; Furlan
\etal 2006) and the necessity of having enough short-wavelength opacity in the
inner disk wall (at the dust destruction radius) to explain the magnitude
of the near-infrared excess emission (e.g., Muzerolle \etal 2003b).

If dust growth and settling were an essential part of limiting the MRI, one might
expect accretion rates to increase with age, whereas there is some evidence
that the opposite occurs (Hartmann \etal 1998).  Also, there is no evidence
from the Taurus IRS survey that the disks with the least flaring indicated in
their SEDs, which also tend to have the weakest silicate features suggesting grain
growth, have larger mass accretion rates than those stars with less evidence
for flaring and grain growth (Furlan \etal 2005, 2006).  
On the other hand, Fromang \etal (2002) took a more optimistic stance, suggesting
that relatively small fractions of metal ions left out of grains could produce
sufficient ionization.  However, even in this case,
Fromang \etal note that when $\alpha \leq 10^{-3}$
dead zones tend to appear no matter what the metal ion fraction becomes, simply
because the X-rays do not penetrate through the entire disk at all radii.

In summary, fully viscous behavior at low disk and stellar masses seems possible
and could be an explanation of very low accretion rates frequently seen in young
brown dwarfs.  The applicability of fully viscous evolution at larger stellar
masses is far more uncertain.

\section{Reynolds stress?}

Fleming \& Stone (2002) conducted numerical simulations of vertically-stratified
disk models in which the upper layers were MRI-active while the central regions
were quiescent.  They found that, although the MRI did not operate in the
magnetically dead zone, the turbulence in the upper layers could generate a significant
Reynolds stress in the midplane, which would allow the dead layer to accrete
albeit with a lower effective viscosity.  

Fleming \& Stone found that the Reynolds stress in the midplane never dropped below
about 10\% of the Maxwell stress in the active layers, and that there was significant
mass mixing between active and dead layers.  If this were generally the case for
protostellar disks, it would simply mean that the effective $\alpha$ parameter
for the evolution of the total surface density would simply be at most an order
of magnitude lower in regions of dead zones.  This would result in an increase
in the mass surface density, which will be inversely proportional to $\alpha$
in steady state (e.g., Reyes-Ruiz \& Stepinski 1995); it would yield a different
dependence of mass accretion rate on disk parameters than in the original layered
model, at large times being dependent on the mass supply from the outer disk
as in the models of Hartmann \etal (1998); and it would limit the buildup
of mass in the dead zone with time in the layered model, eventually
evolving to a steady state.

However, for numerical reasons 
Fleming \& Stone were able only to consider models in which the
surface density of the active layer was of order 20\% of the surface density
of the dead layer or larger.  An active layer of $\sim 100 {\rm g \, cm^{-2}}$
could be 1\% or less of the total surface density at 0.1 AU in
a disk like that of the standard minimum mass solar nebula model.
It seems intuitively unlikely that the turbulence generated in such an
active layer could penetrate so effectively through such a relatively massive
dead zone.  The mass accretion rate depends upon the vertical integral of $\nu \Sigma$;
if the viscosity in the dead zone drops by an order of magnitude, the Reynolds
stresses will have to penetrate to a surface density larger by an order of magnitude or more
in order to have any effect on $\mdot$.  If the Reynolds stresses do not penetrate
to very much deeper layers, they will not change the overall picture of
layered accretion.

\section{Gravitational instability?}

The layered disk model is affected by many aspects of non-thermal ionization
(e.g., grain growth and settling, X-ray fluxes, presence or absence of
cosmic rays) whose effects could range from allowing a full MRI to operate to
shutting it off completely at some radii except for very thin surface layers.
The ubiquity of disk accretion among young stars (when inner disks are present)
exhibiting a substantial but finite range of accretion rates at a given mass
(Hartmann \etal 1998) suggests that some process might be operating which
minimizes the possible variations in MRI activity.

One process of angular momentum transfer that plausibly operates 
during at least the protostellar phase is gravitational instability.
The collapse of the protostellar cloud almost certainly results in most
of the material initially landing at large disk radii, which is a strongly unstable
condition.  It seems highly plausible that the bulk of the protostellar material is
thus driven inward towards the central regions by gravitational torques.  
This raises the question: why not view the T Tauri accretion process as the
not-fully-completed end of the gravitationally-driven phase?  

If gravitational instabilities are dominant in transferring most
of the mass from the initial disk to the central star, the only way for the
disk to be highly gravitationally stable at the end of the protostellar
phase is if the MRI or some other mechanism is very efficient in driving accretion.
While the MRI should be extremely efficient in the inner disk, where thermal ionization
is high, and quite plausibly also in outer disk regions, where X-rays can penetrate
and recombination rates are low, there might still be a dead zone
at intermediate radii which might be driven into gravitational
instability by continually accumulating material from the outer disk.

A major reason why gravitational instabilities have not been strongly advocated
for T Tauri disks is that disk masses seem to be too low.  
Disk masses need to be of the order of 0.1 of the stellar mass for gravitational
instabilities to operate (Pringle 1981), while typical T Tauri disk
mass estimates have been of the order of 0.01 $\msun$, roughly a minimum-mass
solar nebula.  A very recent comprehensive survey of the Taurus star-forming
region suggests a median disk mass of $5 \times 10^{-3} \msun$ (Andrews \&
Williams 2005), albeit with a large scatter.  This median mass estimate
is an order of magnitude lower than what is required for gravitational instability,
though estimates for a small number of objects are much closer to the limiting value.

However, the uncertainty in disk masses estimated from
dust emission is often not sufficiently recognized.
Adoption of typical estimated long-wavelength opacities of interstellar dust 
results in implausibly large disk masses, and the spectral indices of observed
mm- and sub-mm wave emission from T Tauri disks also suggest that the dust 
grains have evolved substantially in the disk (e.g., Beckwith \& Sargent 1991; Andrews \&
Williams 2005, and references therein).  D'Alessio \etal (2001) considered
power-law distributions of dust sizes and showed that dust growth to maximum sizes
greater than about 1 mm results in a spectral index determined by the power law
of the size distribution, not the maximum size of the grains.  
If the maximum size is much larger than 1 mm (for the dust properties
considered by D'Alessio \etal 2001), the mm-wave opacity can be much 
lower than assumed in typical observational estimates, while maintaining a fixed
spectral index determined by the distribution of dust sizes.  In the D'Alessio \etal
calculations, only maximum dust sizes within a factor of 3 to 10 of 1 mm provide
mm-wave opacities close to the standard value assumed; much smaller or much larger
maximum sizes provide much lower opacities.  

Thus, dust growth is more likely to yield low dust opacities than high dust opacities.
Given that some grain evolution
probably has occurred, it seems implausible that it always (or at all radii)
happened to stop at maximum sizes within an order of magnitude of 1 mm.  This
view would suggest that T Tauri disk masses have been systematically underestimated.
It should also be emphasized that the region of the dead zone at $\sim 10$~AU 
may well be optically thick at submm and mm wavelengths (e.g., Beckwith \etal 1990),
so observations of disk dust emission may not be sensitive to the mass in this region.

The typical median dust disk mass estimate is close to the so-called minimum
mass solar nebula (MMSN), $\sim 10^{-2} \msun$, which is basically sufficient to make
Jupiter and little more.  However, a significant number of exoplanets have now been
discovered which have $M \sin i$ larger than one Jupiter mass, sometimes several
Jupiter masses.  It is difficult to believe that such systems can be formed without
having an initial disk mass considerably larger than the MMSN.  While this argument
does not by itself require disk masses large enough to be gravitationally unstable,
it does suggest that disk mass estimates may be systematically low.  

Hartmann \etal (1998) pointed out that accretion rates provide a statistical
constraint on T Tauri disk masses.  For typical estimates of CTTS accretion rates 
$\sim 10^{-8} \msunyr$, lifetimes of 1 - 2 Myr require {\em minimum} disk masses
of $0.01 - 0.02 \msun$, a factor of two to four larger than the median mass estimated
from dust emission by Andrews \& Williams (2005) as discussed above.  There are significant
uncertainties in estimates of accretion rates, which are mostly derived from
ultraviolet and blue-optical continuum excesses, and therefore are sensitive to
extinction corrections to accretion luminosities among other problems.  
On the other hand, it is possible
that the accretion rates estimated by Gullbring \etal (1998), Hartmann \etal (1998),  
and Calvet \& Gullbring (1998) have been underestimated systematically
by not accounting for red-optical excess emission, perhaps by a factor of two or
so (White \& Hillenbrand 2004).  Increasing the accretion rate estimates systematically
would also increase the discrepancy with median dust disk mass estimates.

The solid line in Figure 1 denotes the accretion rate as a function of mass at which
0.1 of the stellar mass would be accreted in $10^6$~yr.  Accretion rates above
this line, if sustained over 1 Myr, would result in accretion of more than $0.1 M_*$
and thus would imply that the disk would have been gravitationally unstable.
The steep dependence of accretion rate on mass means that
the highest-mass pre-main sequence stars in this diagram -- the intermediate-mass
($\sim 1 - 2.5 \msun$) classical T Tauri stars (Calvet \etal 2004) -- 
come much closer to this limiting line than the lower-mass objects.
Given the obvious point that the accretion to date has not
exhausted disk masses, the accretion rate behavior shown in Figure 1 
suggests that gravitational instability is a real possibility for 
at least the higher-mass stars. 
  
Along these lines, it is worth considering accretion in the intermediate-mass stars
closer to the main sequence, the Herbig Ae/Be stars.  It is not clear that X-ray
emission in these stars is sufficient to drive the MRI in their disks.  For example,
the X-ray emission from one of the nearest and best-studied Herbig Ae stars
with an accretion disk, AB Aur, is uncertain.
Zinnecker \& Prebisch (1994) found that AB Aur was a low-luminosity 
($L_X \sim 3 \times 10^{29} {\rm erg s^{-1}}$) X-ray source, despite not
being detected by EXOSAT or EINSTEIN previously; but Damiani \etal (1994)
estimated an upper limit of $L_X < 2 \times 10^{29} {\rm erg s^{-1}}$
from the full set of EINSTEIN IPC observations.  More generally,
by using deep Chandra observations of the very young
Orion Nebula Cluster, Stelzer \etal (2005) showed that of 11 mid B- to late-A
stars, four were not detected with X-ray luminosity upper limits much lower 
than that of the late-type stars in the region, and argued that their results
were consistent with a scenario in which the X-rays in the detected objects
are is dominated by emission from an unresolved late-type companion star.

Even if Herbig Ae/Be stars do have intrinsic X-ray emission at levels comparable
to their lower-mass T Tauri counterparts, higher accretion rates $\sim 10^{-7} \msunyr$
(Muzerolle \etal 2004) may be difficult to sustain.
Interestingly, Grady \etal (1999), Fukagawa \etal (2004), 
Corder, Eisner, \& Sargent (2005), and Pietu \etal (2005) have found evidence for spiral density
wave structure in the outer disk of AB Aur.  This suggests that either that gravitational
instabilities are present, or that the disk being perturbed by a massive object
suggesting gravitational fragmentation; either case is consistent with a large 
outer disk mass, at least initially.

Although gravitational torques can transfer angular momentum outward in disks
and thus drive accretion
(e.g., Tomley \etal 1991; Laughlin \& Bodenheimer 1994; Laughlin \etal 1998),
substantial uncertainties remain, including whether the disk fragments or not
(Johnson \& Gammie 2005).  Models have been constructed in which gravitational
instabilities resulting from pile-up of material in a dead zone can result
in triggering limit-cycle accretion behavior to explain FU Ori outbursts
(Armitage, Livio, \& Pringle 2001; Book \& Hartmann 2005;
Book, Gammie, \& Hartmann, in preparation).
Thus it is not clear that gravitational torques will result in a quasi-steady 
accretion rate at late evolutionary stages.

The main reason for considering gravitational torques is to limit the sensitivity
of accretion in the layered model to dust growth, settling, and overall depletion.
To the extent that the MRI-activated layer is not sufficiently thick to be able to pass through 
accreting material from outer disk regions, low-level gravitational torques could
help maintain accretion rates onto the central star.

\section{Summary and suggestions for future work}

Revisiting Figure 1, one sees that although a dependence of $dM/dt \propto M_*^2$
accounts for the bulk of the data, it also seems that a relation
$dM/dt \propto M_*$ might account for the upper envelope of accretion rates.
This suggests that objects near the upper limit might be more liable to have
layered accretion and dead zones, while the objects (mostly young
brown dwarfs) falling well below the upper envelope might have much 
lower disk masses and thus be fully magnetically active 
(and have evolved viscously by a substantial amount).
In this view it would be misleading to interpret the observations 
in terms of a universal $dM/dt \propto M_*^2$ relation, as differing processes 
could be dominant in differing stellar mass regimes.

In general it would not be surprising if the lowest mass stars and brown dwarfs had
the lowest mass disks, and thus it seems possible that even modest
X-ray emission could make such low mass disks entirely active, increasing the amount
of viscous evolution and draining the disk onto the central mass most rapidly
to yield low accretion rates at ages of about 1 Myr.  More rapid viscous evolution
means that the accretion rates of the youngest brown dwarfs must be much higher,
by a significant factor.  Thus this model predicts that the decline of mass accretion
rates with age should be faster in brown dwarfs than in the nearly solar-mass
T Tauri stars (Hartmann \etal 1998).  In addition, the brown dwarfs with the very low
mass accretion rates should have very low mm-wave emission indicating (roughly)
very low-mass disks, although the faintness of the emission will make this prediction
difficult to test. 

Observations so far indicate no reason why young brown dwarfs accrete in a qualitatively
different manner from that characteristic of T Tauri stars.  Thus, some of these issues
can be explored by studies of very low mass T Tauri stars, which will be brighter and
easier to study.  The predictions of the previous paragraph can be tested by
obtaining much larger samples of mm-wave emission and accretion rate estimates for
$0.3 - 0.1 \msun$ stars.  If viscous evolution is faster in lower-mass stars, one
should observe a steeper decline of mass accretion rate with age than shown in
the higher-mass samples discussed by Hartmann \etal (1998), and mm-wave emission should
correlate strongly with accretion rate; the current data are not conclusive on this
point, given the limited number of detected brown dwarf disks (Klein \etal 2003;
Scholz \etal 2006), the uncertainties in relating this emission to disk mass,
and uncertainties in the mass accretion rate determinations.

Layered accretion with dead zones is more likely to occur in higher-mass systems.  
The existence of dead (or nearly)
dead zones is important for planet formation, especially in the 1-10 AU radial range;
the higher surface densities are likely to yield faster dust grain growth and settling
to the midplane, and the change in the distribution of
disk mass as a function of radius could slow down the so-called Type II viscous migration
of gap-opening planets (e.g., Lin \& Papaloizou 1986).  
Finally, the possibility of marginally gravitationally-unstable
T Tauri disks should not be discounted given the uncertainties in dust opacities,
at least for the most massive systems. 

The research of L.H. and N.C. was supported in part by NASA grants NAG5-9670,
NAG5-13210, NAG5-10545, and grant AR-09524.01-A from the Space Telescope 
Science Institute.  
PD acknowledges grants from Papiit/UNAM and CONACyT, M\'exico.
Support for this work was also provided by NASA through Contract Number 1257184 issued 
by JPL/Caltech and through the Spitzer Fellowship Program under award 011 808-001.

\facility{}

\begin{figure}
\epsscale{0.7}
\plotone{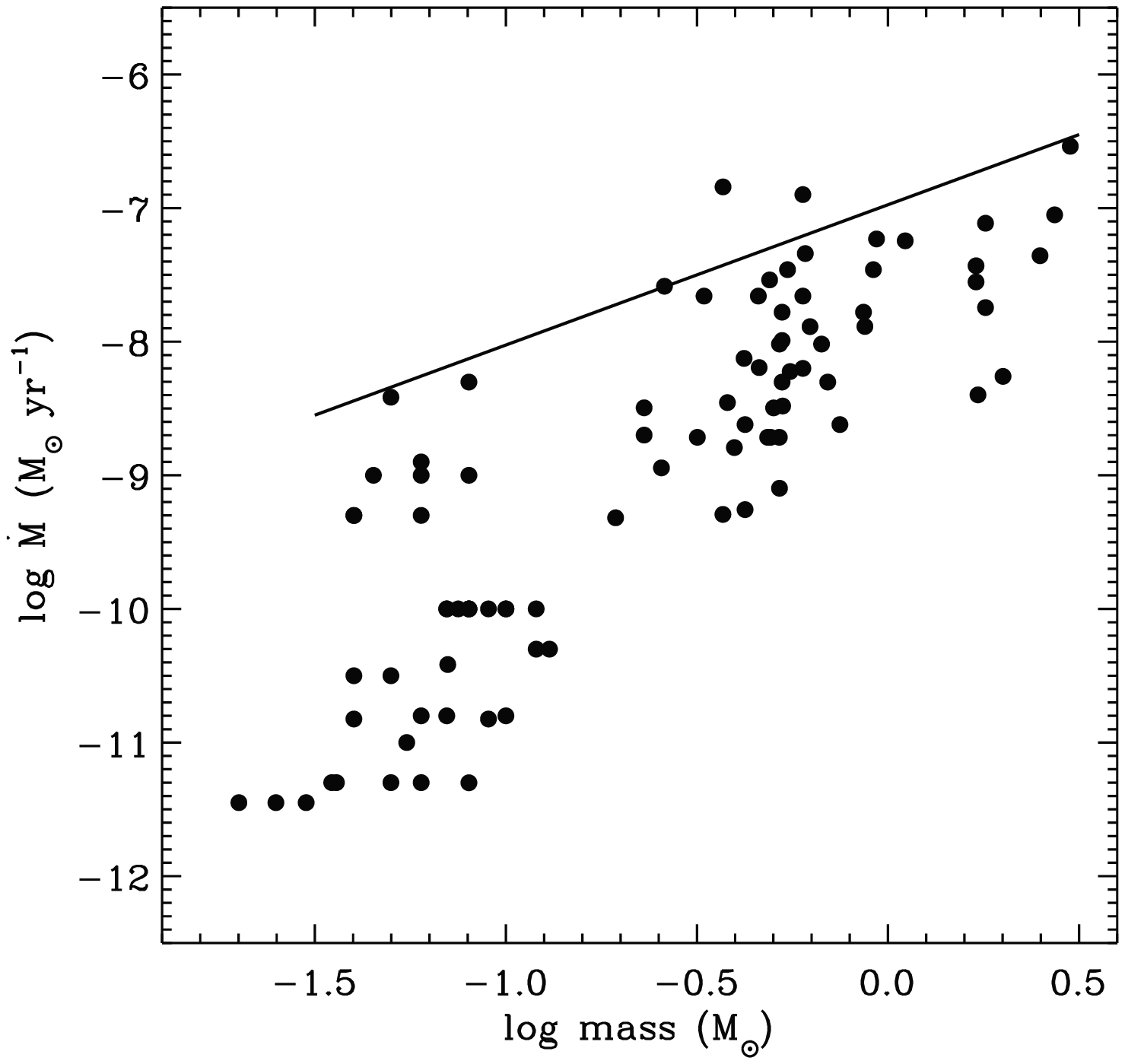}
\caption{Relation of mass accretion rate and stellar mass, reproduced
from Muzerolle \etal (2005), with the addition of a straight line indicating
the relation $\mdot = 0.1 M_*/10^6 {\rm yr}$.  Above this line, sustained
accretion for a typical age of 1 Myr would imply an initial disk mass
likely to be gravitationally unstable (see text)}
\end{figure}

\end{document}